\documentclass[aps,prl,reprint,floatfix,groupedaddress,superscriptaddress,longbibliography]{revtex4-1}
\usepackage{lipsum}
\usepackage{graphicx}
\usepackage[utf8]{inputenc}
\usepackage{amsmath}
\usepackage{empheq}
\usepackage{geometry}
\usepackage{amssymb}
\usepackage{textcomp}
\usepackage{gensymb}

\DeclareGraphicsExtensions{.pdf,.png}

\begin{document}

\newcommand{\vect}[1]{{\bf{#1}}}
\newcommand{\matr}[1]{\mathsf{#1}}
\newcommand{\ymgo}{YbMgGaO$_4$}
\newcommand{\ybt}{Yb$^{3+}$}
\newcommand{\cs}{\left(\frac{d\sigma}{d\Omega}\right)}

\title{Strong exchange anisotropy in \ymgo\ from polarized neutron diffraction}

\author{Sándor Tóth}
\email{sandor.toth@psi.ch}
\affiliation{Laboratory for Neutron Scattering and Imaging, Paul Scherrer Institute, 5232 Villigen PSI, Switzerland}

\author{Katharina Rolfs}
\affiliation{Laboratory for Scientific Developments and Novel Materials, Paul Scherrer Institute, 5232 Villigen PSI, Switzerland}

\author{Andrew R. Wildes}
\affiliation{Institut Max von Laue-Paul Langevin, 38042 Grenoble 9, France}

\author{Christian Rüegg}
\affiliation{Laboratory for Neutron Scattering and Imaging, Paul Scherrer Institute, 5232 Villigen PSI, Switzerland}
\affiliation{Department of Quantum Matter Physics, University of Geneva, 1211 Genève, Switzerland}

\date{\textrm{\today}}
%\pacs{75.30.Et,63.20.dd,63.20.kk}
% 75.10.Hk Classical spin models
% 75.30.Ds spin-waves
% 75.30.Et Exchange and superexchange interactions
% 63.20.dd Phonons in crystal lattices: Measurements
% 63.20.kk Phonons in crystal lattices: Phonon interactions with other quasiparticles

\begin{abstract}
We measured the magnetic correlations in the triangular lattice spin-liquid candidate material \ymgo\ via polarized neutron diffraction. The extracted in-plane and out-of-plane components of the magnetic structure factor show clear anisotropy. We found that short-range correlations persist at the lowest measured temperature of 52 mK and neutron scattering intensity is centered at the $M$ middle-point of the hexagonal Brillouin-zone edge. Moreover, we found pronounced spin anisotropy, with different correlation lengths for the in-plane and out-of-plane spin components. When comparing to a self-consistent Gaussian appoximation, our data clearly support a model with only first-neighbor coupling and strongly anisotropic exchanges. 
\end{abstract}

\maketitle

%%%%%%%%%%%%%%%%%%%%%%%%%%%%%%%%%%%%%%%%%%%%%%%%%%%%%%%%%%%%%%%%%%%%%
%\section{Introduction}
%%%%%%%%%%%%%%%%%%%%%%%%%%%%%%%%%%%%%%%%%%%%%%%%%%%%%%%%%%%%%%%%%%%%%

Anderson proposed in a seminal paper that the ground state of the spin-1/2 Heisenberg triangular lattice antiferromagnet (TLA) is a "quantum liquid" of resonating valence bonds \cite{Anderson1973}. Later studies of this model mostly showed an ordered ground state, with sublattice magnetization $\langle S\rangle=0.41S$ \cite{Gammon1991,Runge1993,Elstner1993,Capriotti1999,White2007,Kulagin2013}. In contrast to the isotropic model, perturbations such as further-neighbor interactions \cite{Manuel1999,Li2015c,Zhu2015,Hu2015,Saadatmand2016,Iqbal2016} or ring exchange \cite{Motrunich2005,Grover2010} were shown to destroy Néel order and to promote the formation of a spin-liquid ground state. Real, undistorted spin-1/2 triangular lattice systems are rare. Until recently the charge transfer salts were the only known class. Members such as $\kappa$-(BEDT-TTF)$_2$Cu$_2$(CN)$_3$ \cite{Yamashita2008} and EtMe$_3$Sb[Pd(dmit)$_2$] \cite{Yamashita2010} show spin-liquid ground states with a spinon Fermi surface at low temperature. However the magnetism of these systems is complicated with charge fluctuations inducing an effective ring exchange between spins that destroys Néel order. Also, no momentum resolved spectroscopic data of the magnetic excitations, crucial to fully characterize the correlated state, is available due to the small size of the synthetic crystals. 

% sample size

\begin{figure}[!htb]
    \centering
	\includegraphics[width=\columnwidth]{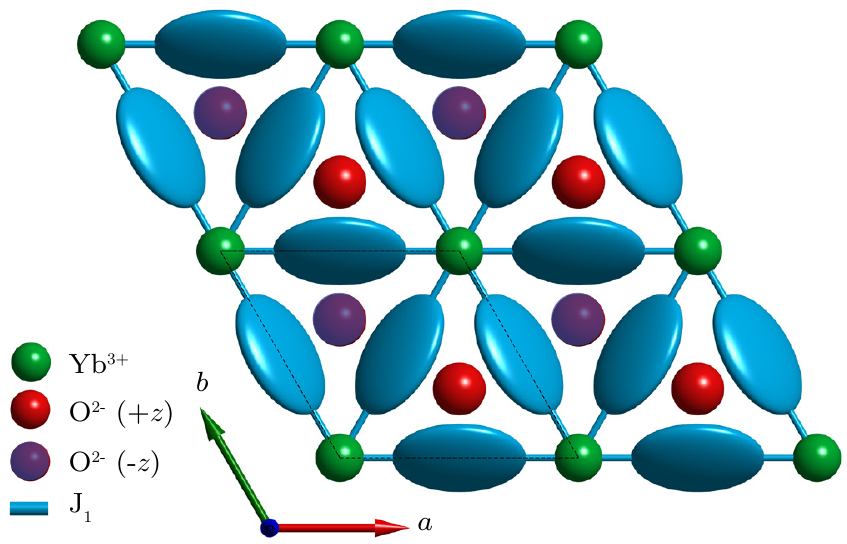}
	\caption{Single triangular layer of \ymgo\ with the magnetic Yb$^{3+}$ atoms shown as green spheres. Surrounding oxygen atoms above and below the triangular plane are shown by red and purple spheres, respectively. The tilted blue ellipsoids represent the first-neighbor symmetric exchange tensor \cite{spinw}.}
	\label{fig:struct}
\end{figure}

The recent discovery of \ymgo, an inorganic rare-earth oxide with effective spin-1/2 triangular lattice (see Fig.\ \ref{fig:struct}) gives a new opportunity to study frustrated magnetism in this simple geometry \cite{Li2015b}. First studies showed correlated spin fluctuations without long range order at the lowest measured temperature of 50 mK \cite{Li2015}. Moreover, large single crystals enabled momentum resolved neutron spectroscopic studies of the magnetic excitation spectrum \cite{Paddison2016} that was interpreted both as a spinon Fermi surface \cite{Shen2016} and resonating valence bonds \cite{Li2017g}. Supporting the resonating valence bond picture, recent  measurements found no magnetic contribution to the thermal conductivity \cite{Xu2016a}. Moreover it has emerged very recently that crystalline electric-field randomness induced by the non-magnetic Mg/Ga site disorder is an essential characteristic of \ymgo \cite{Li2017}. This suggests that the origin of the low temperature disordered state might be driven by both intrinsic (quantum fluctuations) and extrinsic (exchange randomness) parameters. The theoretical development is also hampered by the lack of consensus on the underlying spin Hamiltonian.

In this Letter, we report the results of polarized neutron diffraction measurements in the candidate spin-liquid phase of \ymgo. The experimental technique probes separately the real-space dependence and the spin-direction dependence of the spin-spin equal time correlation function. In rare earth systems such as \ymgo, the magnetic correlations in spin and real space are entangled due to strong the spin orbit coupling \cite{Chun2015}. By measuring the correlations between different spin components, we could sensitively measure the anisotropic components of the exchange interactions that are important stabilizing the spin-liquid ground-state. We could unambiguously identify the underlying spin Hamiltonian by comparing our data to self-consistent Gaussian approximation (SCGA).

We synthesized single-crystal samples of \ymgo\ using the floating-zone method, described elsewhere \cite{Li2015}. The magnetic susceptibility is identical to previous reports \cite{Li2015b,Li2015}. The data show no sign of magnetic order down to 0.4 K and the previously reported weak plateau-like feature in the out-of-plane magnetization around 2 T at 0.4 K, corresponding to half the saturation magnetization is also observed. We measured polarized neutron diffraction on the D7 instrument at ILL, France \cite{Stewart2009}. The sample consisted of three coaligned single crystals with a total mass of 1.12 g and a mosaicity of 4\degree. We selected an incident neutron wavelength of 4.86 \AA\ (3.47 meV) where the incident neutron flux is highest. The horizontal scattering plane was the triangular plane ($ab$) and the sample was rotated around the vertical axis, while data were recorded. To separate the magnetic signal from the instrument background, nuclear-coherent and nuclear-spin-incoherent scattering, we employed $xyz$-polarization analysis \cite{Stewart2009}. We collected data for all three neutron polarization directions and for both spin-flip (SF) and non-spin-flip (NSF) channels. 

\begin{figure}[!htb] 
    \centering
	\includegraphics[width=\columnwidth]{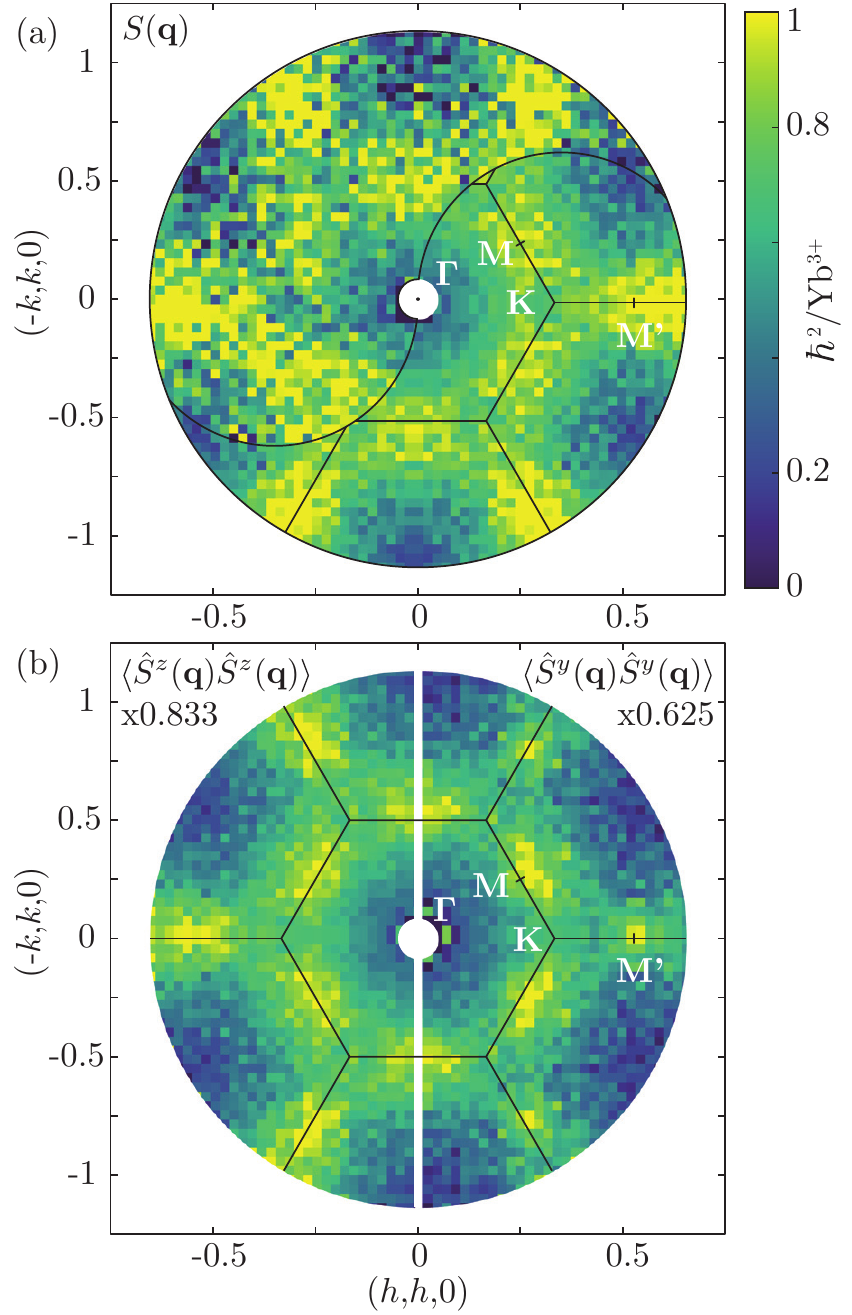}
	\caption{Map of the magnetic diffuse scattering signal in \ymgo, measured at $T=52(2)$ mK. (a) Magnetic scattering, the upper left region shows the measured data, while the lower right region is symmetrized. (b) The $S^{yy}(\vect{q})$ in-plane and $S^{zz}(\vect{q})$ out-of-plane components of the spin-spin equal time correlation function, the corresponding SCGA calculation is shown in Fig.\ S1 in the Supplementary Materials. Note that intensity is scaled to the maximum signal on both sides.}
	\label{fig:diff}
\end{figure}

The diffuse magnetic scattering signal of \ymgo, measured at 52(2) mK, is shown in Fig. \ref{fig:diff}(a). The data is calculated from a linear combination of the six measured neutron polarization channels, see Sec.\ 2 of the Supplementary Materials \cite{supp}. The resulting signal is purely magnetic, free of background and it is in absolute units. The measured magnetic signal shows a clear momentum dependence, with intensity concentrated around the equivalent $M$- and $M'$-points of the Brillouin-zone (BZ). Since the incident neutron energy is much higher than the 1.5 meV bandwidth of the magnetic excitation \cite{Shen2016}, we can apply the static approximation \cite{Squires}. In this case the measured signal $S(\mathbf{q})$ corresponds to the linear combination of equal-time magnetic correlation functions $S^{\alpha\beta}(\mathbf{q})$ defined as
\begin{align}
S(\mathbf{q}) &= \sum_\alpha\int_0^\infty (1-\hat{q}_\alpha^2) g^2_\alpha S^{\alpha\alpha}(\mathbf{q},\omega)d\omega\nonumber\\
			  &=\sum_\alpha(1-\hat{q}_\alpha^2)g_\alpha^2S^{\alpha\alpha}(\vect{q}),
\end{align}
where $g_\alpha$ denotes the diagonal values of the g-tensor ($g_x=g_y=3.06$, $g_z=3.721$ \cite{Li2015}).

The $xyz$ neutron polarization analysis also enables the separation of the $S^{yy}(\mathbf{q})=\langle \hat{S}(\mathbf{q})^y\hat{S}(-\mathbf{q})^y\rangle$ and $S^{zz}(\mathbf{q})=\langle \hat{S}(\mathbf{q})^z\hat{S}(-\mathbf{q})^z\rangle$ components of the spin-spin correlation function. Here the $y$ axis is in the triangular plane and always perpendicular to the momentum vector $\vect{q}$ and $z$ is vertical. The two components are shown in Fig.\ \ref{fig:diff}(b) revealing a difference. The $S^{zz}$ component peaks more sharply at the $M$ and $M'$ points and the overall strength of the correlations is larger (note that intensity map is scaled independently to the maximum signal of $S^{yy}$ and $S^{zz}$). 

\begin{figure}[!htb]
    \centering
	\includegraphics[width=\columnwidth]{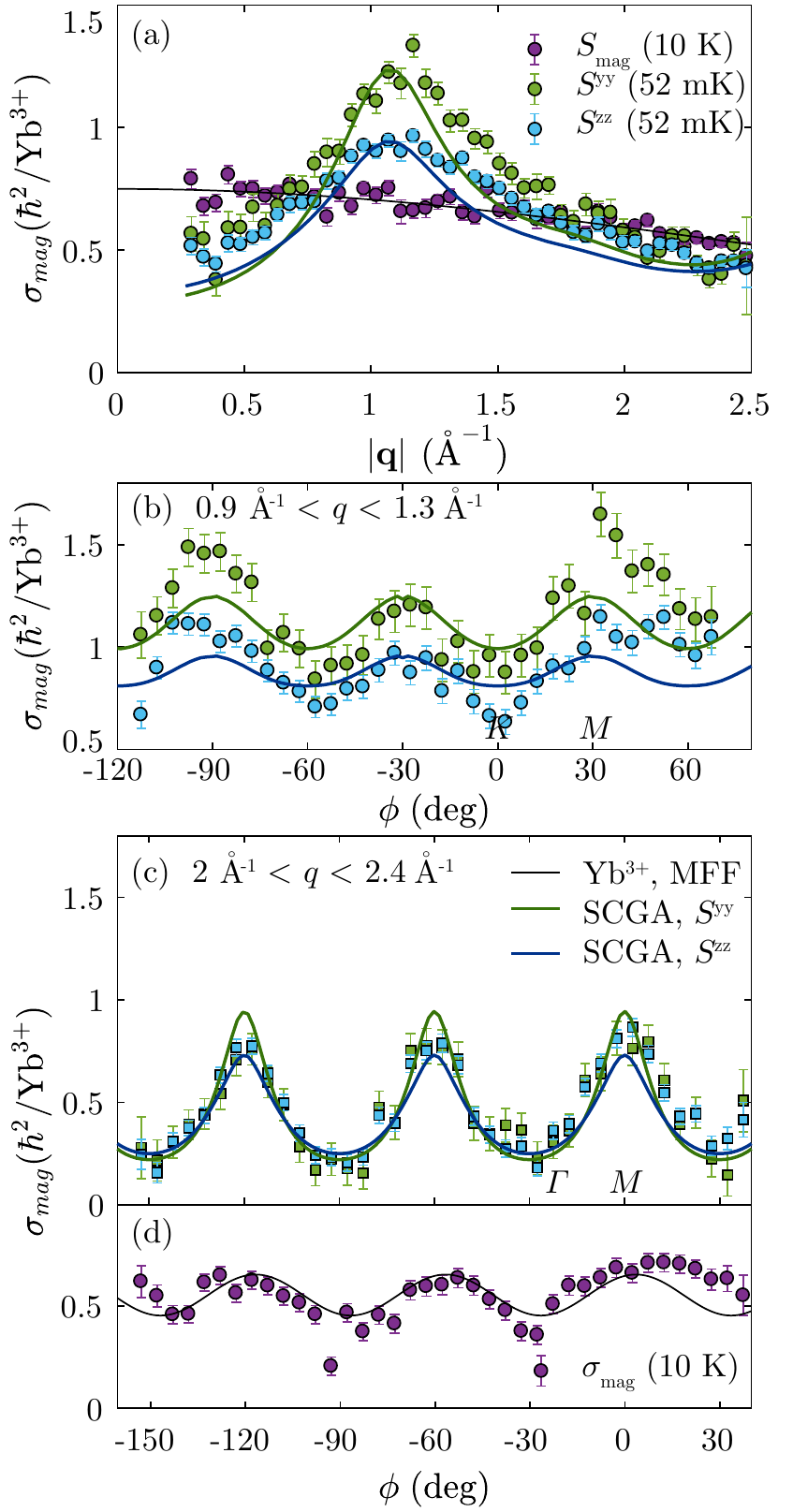}
	\caption{Integrated magnetic diffuse scattering signal of \ymgo\ measured at different temperatures and compared to the SCGA calculation. Purple symbols denote the magnetic signal measured at 10 K, green and blue symbols correspond to the measured spin structure-factors $S^{yy}$ and $S^{zz}$ at 52(2) mK, while the blue and green lines show the corresponding SCGA result simulated at $T_{MF}=120$ mK for model B. (a) Powder-averaged data  with black line denoting the magnetic form factor of the Yb$^{3+}$ ion. (b-d) Angular dependence of the diffuse scattering signal integrated between $q=0.9-1.3$ \AA$^{-1}$ and $q=2-2.4$ \AA$^{-1}$. Black line is a guide to the eye.}
	\label{fig3}
\end{figure}

The spin anisotropy of the correlations in \ymgo\ is best visible in line plots of the non-symmetrized data shown in Fig.\ \ref{fig3}. It is informative to calculate the powder-averaged magnetic correlations as a function of $q=|\vect{q}|$ shown in Fig.\ \ref{fig3}(a). The signal, measured at 10 K, is identical to the magnetic form factor of Yb$^{3+}$, confirming the ideal paramagnetic nature of \ymgo\ at this temperature. At base temperature the correlations between the in-plane spin components has a more pronounced maximum when compared to the $z$-component and both are centered at $q=1.15$ \AA$^{-1}$. We also integrated the magnetic signal within the $0.9<q<1.3$ \AA$^{-1}$ annulus that includes the $M$- and $K$-points of the first BZ, see Fig.\ \ref{fig3}(b). Both the $y$ and $z$ components of the effective spin show clear peaks at the symmetry equivalent $M$-points of the first BZ. We also integrated the magnetic signal between $2.0<q<2.4$ \AA$^{-1}$ that includes the center of the second BZ, see Fig.\ \ref{fig3}(c). The signal again shows clear peaks at angles corresponding to the $M'$-points while there is no difference between the $y$ and $z$ components of the spin in this cut.

%%%%%%%%%%%%%%%%%%%%%%%%%%%%%%%%%%%%%%%%%%%%%%%%%%%%%%%%%%%%%%%%%%%%%%%%%%%%%%%%%%%%
% DISCUSSION
%%%%%%%%%%%%%%%%%%%%%%%%%%%%%%%%%%%%%%%%%%%%%%%%%%%%%%%%%%%%%%%%%%%%%%%%%%%%%%%%%%%%

The momentum dependence of the measured magnetic diffuse scattering in \ymgo\ reveals short-range correlations down to the lowest temperature of 52 mK. This temperature is far below the Curie-Weiss temperature of 4 K \cite{Li2015b}, suggesting a strongly fluctuating magnetic ground state. The most prominent feature of the diffuse scattering data is the peak at  $(1/2,0,0)$, the $M$ point of the Brillouin zone as previously observed via neutron scattering without separating the spin components \cite{Paddison2016}. This is in strong contrast to the results of the simplest antiferromagnetic model on the triangular lattice with isotropic first-neighbor interactions, where correlations are strongest at the $K$-point $(1/3,1/3,0)$. This suggests that further terms in the Hamiltonian are important in describing the low temperature correlated state of \ymgo. Our second observation is that the correlations are strongly spin direction dependent, revealing that the underlying Hamiltonian has to be anisotropic as well. This is indeed expected for a system with strong spin-orbit coupling and previously evidenced by magnetic susceptibility, magnetization and electron spin resonance measurements in the paramagnetic phase  \cite{Li2015b,Li2015}. Also it was previously shown that strongly anisotropic couplings are present in the Yb$^{3+}$-pyrochlore compound, Yb$_2$Ti$_2$O$_7$ \cite{Ross2011}.

Two different model Hamiltonians have been proposed that describe the exchange interactions in \ymgo \cite{Paddison2016,Li2016a}. The for the ambiguity of the spin wave fit is two fold. Firstly, the spin waves in the magnetic field polarized phase are broad due to the randomness of the g-values \cite{Li2017}. Secondly, the spin wave spectrum is independent of $J_{z\pm}$ for the measured $B\|z$ field direction. In order to compare these models with our experimental data, we use self-consistent Gaussian approximation to calculate spin-spin correlations \cite{Garanin1996,Bergman2007}, as detailed in Sec.\ 4 of the Supplementary Materials \cite{supp}. The SCGA method describes correlations between classical spins taking into account the fluctuations of the molecular field. It is accurate except the vicinity of the critical point $T_c$. All model parameters are fixed by previous fits of the spin wave dispersion except the model temperature, that is fitted to our diffuse scattering data. The reason for taking temperature as a variable is that we substitute quantum fluctuations with thermal fluctuations in the model.

The general model Hamiltonian compatible with the $R\overline{3}m$ crystal symmetry of \ymgo\ and including anisotropic first and second neighbor interactions is given by
\begin{align}
	\mathcal{H} =& \sum_{\langle ij \rangle} [J_{zz}S_I^zS_j^z +J_\pm(S_i^+S_j^-+S_i^-S_j^+)\\
				 &+J_{\pm\pm}(\gamma_{ij}S_I^+S_j^++\gamma^*_{ij}S_i^-S_j^-)\nonumber\\
				 &-\frac{iJ_{z\pm}}{2}(\gamma^*S_i^+S_j^z-\gamma_{ij}S_i^-S_j^z+\langle i \leftrightarrow j\rangle)]\nonumber\\
				 &+\sum_{\langle\langle ij\rangle\rangle} [J_{2z}S_i^zS_j^z+J_{2\pm}(S_i^+S_j^- + S_i^-S_j^+)],\nonumber
\end{align}
where $S_i^\pm=S_i^x\pm S_i^y$, and the phase factor $\gamma_{ij}=1,\, e^{i2\pi/3},\, e^{-i2\pi/3}$. The antisymmetric terms (Dzyaloshinskii-Moriya interactions) are forbidden by the lattice symmetry. The first model (model A) includes first- and second-neighbor exchange interactions in the triangular plane and assumed $J_{z\pm}=0$ ($J_{zz}=126$ $\mu$eV, $J_{\pm\pm}=13$ $\mu$eV, $J_\pm=109$ $\mu$eV, $J_{2z}=27$ $\mu$eV,  with all other terms set to zero) \cite{Paddison2016}. The second model (Model B) assumes only first-neighbor interactions with the additional anisotropic exchange term ($J_{zz}=164$ $\mu$eV, $J_{\pm\pm}=56$ $\mu$eV, $J_\pm=108$, $J_{z\pm}=98$ $\mu$eV) \cite{Li2016a}. Both models produce a peak in the structure-factor at the $M$-point and fits the field polarized spin wave spectrum.

Our SCGA solution of the two models show, that the difference between the $S^{yy}$ and $S^{zz}$ correlations is sensitive to the exchange anisotropy, thus providing additional crucial information about the magnetic Hamiltonian. Using an effective temperature of 120 mK and the parameters of model B, the calculation reproduces the low-temperature correlations of both the $y$- and $z$-components of the effective spins very well, as shown in Figs.\ \ref{fig3}(a-c) (and in Fig.\ S1 in Supplementary Materials). Model A has a clearly worse agreement with the data, see Fig.\ S1 in Supplementary Materials. Model B accounts for both the spin anisotropy in Figs.\ \ref{fig3}(a-b) and the isotropic character of the higher momentum transfer cut in Fig.\ \ref{fig3}(c). The simulation also reproduces the absolute intensities, suggesting that the data contains no background signal. The SCGA calculation also recovers the lack of correlations at 10 K, in agreement with the measured paramagnetic signal (Fig.\ \ref{fig3}(a)). However the measured modulation of the structure-factor in Fig.\ \ref{fig3}(d) is not reflected in the calculation, which gives a completely flat signal. Thus the angular modulation of the structure-factor at 10 K is the property of the magnetic Yb$^{3+}$ ions. A modulated signal in the paramagnetic phase is possible if the magnetic form factor of Yb$^{3+}$ is slightly anisotropic, as expected for Yb$^{3+}$ ions in the local $D_{3d}$ point group symmetry \cite{Rotter2009}.

% DISCUSSION

Our results shows that the effective spin-1/2 Hamiltonian of \ymgo\ contains only first neighbor interactions. This is in agreement with the strongly localized nature of the $4f$ electrons on Yb$^{3+}$. Moreover, second neighbor superexchange interactions require electrons hopping over two oxygen atoms located on the opposite sides of the magnetic triangular planes, see Fig. \ref{fig:struct}. Additional dipolar interactions are negligible due to the large distance. Also a well studied related compound, Yb$_2$Ti$_2$O$_7$ with similar bond length have negligible second neighbor exchange interactions \cite{Ross2011}. Although the exchange values determined from magnetization and ESR line-width \cite{Li2015} measurements seem to deviate from model B, in the analysis of the ESR data the recently found g-tensor randomness was not taken into account \cite{Li2017}. The observed broad distribution of the g-values have strong influence on the field dependent ESR spectrum.

The best fitting model also suggests, that the theoretical search for the spin liquid or resonating valence bond ground state should be extended towards more generic Hamiltonians that include terms beyond the recently investigated pseudo dipolar ($J_{\pm\pm}$) interaction \cite{Li2017}. The recently found crystalline electric-field randomness \cite{Li2017} has led to the proposal that the randomness of the pseudo-dipolar term can induce a disordered ground state, which can be identified as a spin-glass like state \cite{Zhu2017}. In this work the XXZ model with additional pseudo-dipolar interactions was studied via DMRG calculations showing that the parameters of model A put the system deep into the stripe ordered phase. The analysis attributed the experimentally observed lack of long range order to the Ga/Mg site disorder by introducing spatially random values of $J_{\pm\pm}$. Although the existence of site disorder and g-tensor randomness was shown experimentally \cite{Li2017}, there is no evidence of the proposed exchange randomness. According to our analysis, it is absolutely necessary to extend the theoretical search to more general first neighbor Hamiltonians and potentially including exchange disorder and to disentangle the role of the intrinsic (quantum fluctuations) and extrinsic (site disorder) parameters in the observed low temperature magnetic phase. First study in this direction showed that the stability of the 120\degree\ magnetic order in the XXZ model is decreased due to enhanced quantum fluctuations when both the anisotropic $J_{\pm\pm}$ and $J_{z\pm}$ interactions are included \cite{Liu2016,Li2016d}.

%%%%%%%%%%%%%%%%%%%%%%%%%%%%%%%%%%%%%%%%%%%%%%%%%%%%%%%%%%%%%%%%%%%%%
% CONCLUSION
%%%%%%%%%%%%%%%%%%%%%%%%%%%%%%%%%%%%%%%%%%%%%%%%%%%%%%%%%%%%%%%%%%%%%

In conclusion, we showed that the equal-time spin-spin correlations of \ymgo\ vary in a manner consistent with anisotropic exchange interactions between first neighbor pseudospins. By comparing the measured spin structure-factor to the SCGA calculation of previously proposed models, we could show that the first-neighbor-only model describes our experimental data best. Our confirmation of the spin Hamiltonian of \ymgo\ shows the direction where theory should search for the description of the proposed spin-liquid phase.

\begin{acknowledgements}
	The authors would like to thank Bruce Normand and Tom Fennell for useful discussion and Lucile Mangin-Thro for the help at the D7 experiment. This research was supported by the SNF Sinergia Project "Mott Physics beyond Heisenberg".
\end{acknowledgements}

\bibliographystyle{apsrev4-1}
%\bibliography{/Users/sandortoth/Papers/library}

\end{document}